\documentclass[aps,prd,twocolumn,showpacs,nofootinbib]{revtex4-1}

\usepackage{graphicx}
\usepackage{amsmath}
\usepackage{amssymb}
\usepackage{bm}
\usepackage{bbm}
\usepackage{color}
\usepackage{slashed}
\usepackage[colorlinks=true,linkcolor=blue,citecolor=blue, urlcolor=blue]{hyperref}

\begin{document}

\title{Higgs boson decays to $B_c$ meson in the fragmentation-function approach}

\author{Xu-Chang Zheng}
\email{zhengxc@cqu.edu.cn}
\author{Xing-Gang Wu}
\email{wuxg@cqu.edu.cn}
\author{Xi-Jie Zhan}
\email{zhanxj@cqu.edu.cn}
\author{Guang-Yu Wang}
\email{wanggy@cqu.edu.cn}
\author{Hong-Tai Li}
\email{liht@cqu.edu.cn}

\affiliation{Department of Physics, Chongqing Key Laboratory for Strongly Coupled Physics, Chongqing University, Chongqing 401331, People's Republic of China}

\begin{abstract}

In the paper, we present a calculation of the decay widths for the Higgs boson decays to the $B_c$, $B_c^*$, $B_c(2^1S_0)$ and $B_c^*(2^3S_1)$ mesons using the fragmentation-function approach. In the calculation, the fragmentation functions up to order $\alpha_s^3$ based on the nonrelativistic QCD factorization theory are used, and the decay widths for $H\to Q+X$ and $H \to g+X$ at the partonic level are calculated up to order $\alpha_s$. The large logarithms of $m_H^2/m_{Bc}^2$ are resummed up to next-to-leading logarithmic accuracy by solving the evolution equations for the running quark masses and the fragmentation functions. Compared to the leading-order decay widths based on the nonrelativistic QCD approach, the decay widths based on the fragmentation-function approach that include the higher-order QCD corrections are reduced significantly. Our numerical results show that there are about $1.2\times 10^5$ $B_c$ events via the Higgs decays to be produced at the HL-LHC with $3ab^{-1}$, and about $1.6\times 10^6$ $B_c$ events via the Higgs decays to be produced at the HE-LHC with $15ab^{-1}$.

\end{abstract}


\maketitle

\section{Introduction}

The discovery of the Higgs boson at the LHC \cite{ATLAS:2012yve,CMS:2012qbp} in 2012 is an important breakthrough in our understanding of fundamental interactions. After that, an important task is to accurately study the properties of the Higgs boson, including the Higgs couplings to the fundamental fermions and the gauge bosons, as well as the Higgs self-coupling, and test whether these couplings are completely consistent with those predicted by the Standard Model (SM). Any measurement that deviates from the SM prediction may be a signal of new physics.

The LHC has achieved great success in discovering the Higgs boson, and has studied some couplings of the Higgs boson, e.g., the couplings to heavy vector bosons \cite{ATLAS:2015muc,ATLAS:2018pgp,CMS:2018zzl} and the charged fermions of the third generation \cite{ATLAS:2016neq,ATLAS:2018mme,CMS:2018uxb,ATLAS:2018kot,CMS:2018nsn,ATLAS:2019nkf,CMS:2018uag,ATLAS:2019nkf,Noguchi:2019ofj,CMS:2017zyp}. However, the precision of these measurements is restricted due to the limited Higgs events and complicated hadronic background. After a period of shut down, the LHC has just upgraded to Run 3. During Run 3, more data will be collected than the first two runs combined. Futhermore, the LHC is planned to upgrade to the High-luminosity LHC (HL-LHC) and the High-energy LHC (HE-LHC) after Run 3. At the HL-LHC ($\sqrt{s}=14\,{\rm TeV}$), with an integrated luminosity of $3\,ab^{-1}$, about $1.6 \times 10^8$ Higgs boson events will be produced; At the HE-LHC ($\sqrt{s}=27\,{\rm TeV}$), with an integrated luminosity of $15\,ab^{-1}$, about $2.2 \times 10^9$ Higgs boson events will be produced \cite{Cepeda:2019klc}. In addition, several lepton colliders are under consideration, e.g., the Circular Electron-Positron Collider (CEPC) \cite{CEPCStudyGroup:2018ghi}, the International Linear Collider (ILC) \cite{ILC:2013jhg}, and the $e^+e^-$ Future Circular Collider (FCC-ee) \cite{FCC:2018evy}, and the Muon Collider \cite{MuonCollider:2022xlm,Black:2022cth}. One of the advantages of lepton colliders is that the background is clean, thus they are suitable for the precision measurements of the properties of the Higgs boson.

With these collider platforms, some rare decays of the Higgs boson, such as the Higgs decays to quarkonium, may be measured \cite{Qiao:1998kv,Bodwin:2013gca,Bodwin:2014bpa,Konig:2015qat,Zhou:2016sot,Modak:2016cdm,Bodwin:2016edd,Bodwin:2017wdu,Sun:2018xft,Liao:2018nab,Brambilla:2019fmu,
Mao:2019hgg,Liao:2019xux,Sun:2019cxx,Pan:2022nxc,Han:2022rwq,Gao:2022iam,Batra:2022wsd,Jiang:2015pah}. These rare decays can be used to determine the magnitude of the Yukawa couplings of the Higgs boson to the heavy quarks, and they have distinguished signals to be detected at the high-luminosity or high-energy colliders. The $B_c$ meson carries two different heavy flavors and provides a unique bound-state system for testing the SM. As a combination, in Ref.\cite{Jiang:2015pah}, the authors studied the Higgs decays to the $B_c$ meson at the leading order (LO) accuracy, and they found that about $1.4 \times 10^5$ $B_c$ events can be produced through the Higgs decays at the HL-LHC. In addition to studying the Higgs properties, this decay process can also be used to study the production mechanism of the $B_c$ meson. Thus, it is attractive to present a more precise study on this decay process. In the present paper, we devote ourselves to reanalyzing this decay process with higher accuracy in the fragmentation-function approach.

There are large logarithms of the form ${\rm ln}(m_H^2/m_{Bc}^2)$ in the perturbative series of the decay width of the Higgs boson into a $B_c$ meson, which come from two sources: the renormalization of the Yukawa couplings and the emission of the collinear gluons. These large logarithms may spoil the convergence of the perturbative expansion, thus it is important to sum them to all orders (in $\alpha_s$) in the calculation. It is noted that under the fragmentation-function approach, the large logarithms from these two sources can be resummed simultaneously. More explicitly, the large logarithms from the renormalization of the Yukawa couplings can be resummed by using the running quark masses for the heavy ($b$ and $c$) quarks \cite{Braaten:1980yq,Sakai:1980fa}; while the large logarithms from the collinear gluon emission can be resummed through solving the Dokshitzer-Gribov-Lipatov-Altarelli-Parisi (DGLAP) evolution equations for the fragmentation functions of the $B_c$ production \cite{Zheng:2019egj,Zhang:2021ypo}. In this paper, we will resum these large logarithms up to next-to-leading logarithmic (NLL) accuracy under the fragmentation-function approach.

Because of carrying two different heavy flavors, the excited states of the $B_c$ meson below the BD threshold\footnote{The $B_c$ excited states above the BD threshold will decay mainly into a pair of $B$ and $D$ mesons \cite{Li:2019tbn}.} will decay to the ground state $B_c$ meson with almost $100\%$ probability through electromagnetic or strong interaction. Thus, the excited states are important sources of the ground state $B_c$ meson. Furthermore, the production of these excited states via the Higgs boson decays is also interesting by itself. Therefore, besides the decay width for the Higgs boson into the ground state $B_c$ meson, we will also calculate the decay widths for the Higgs boson decays into the $S$-wave excited states, e.g. $B_c^*$, $B_c(2^1S_0)$, and $B^*_c(2^3S_1)$.

The paper is organized as follows. In Sec.\ref{sec2}, we present useful formulas for the decay width of the Higgs boson into the $B_c$ meson under the fragmentation-function approach. In Sec.\ref{sec3}, numerical results and discussions are presented. Section \ref{sec4} is reserved as a summary.

\section{Calculation formalism}
\label{sec2}

In this section, we present useful formulas for the considered decay width under the fragmentation-function approach. For simplicity, we only give the formulas for the ground state $B_c$ meson, the formulas for the excited states [i.e., $B_c^*$, $B_c(2^1S_0)$, and $B^*_c(2^3S_1)$] are similar to the ground state $B_c$ case.

Under the fragmentation-function approach, the differential decay width for the decay channel $H \to B_c+X$ can be written as
\begin{eqnarray}
\frac{d\Gamma_{H \to B_c+X}}{dz}=&& \sum_{i} \int_z^1 \frac{dy}{y}\frac{d \hat{\Gamma}_{H \to i+X}(y,\mu_F)}{dy}\nonumber \\
&& \times D_{i \to B_c}(z/y,\mu_F)+{\cal O}(m_{B_c}^2/m_H^2),
\label{pQCDfact}
\end{eqnarray}
where $d \hat{\Gamma}_{H \to i+X}(y,\mu_F)/dy$ stands for the differential decay width of $H \to i+X$ at the partonic level, $D_{i \to B_c}(z/y,\mu_F)$ stands for the fragmentation function for a parton $i$ into the $B_c$ meson, $z=2p_{B_c} \cdot P_H /m_H^2$ denotes the energy fraction carried by the $B_c$ meson from the Higgs boson, $\mu_F$ denotes the factorization scale, and the sum extends over the patron species.

The decay widths for the Higgs boson into a parton can be calculated through perturbation theory. Up to now, the decay width for the Higgs boson into bottom quarks has been calculated up to order $\alpha_s^4$ \cite{Braaten:1980yq,Sakai:1980fa,Gorishnii:1990zu,Kataev:1993be,Surguladze:1994gc,Larin:1995sq,Chetyrkin:1995pd,Chetyrkin:1996sr,Baikov:2005rw,Mondini:2019gid}. However, the expressions for the differential decay widths $d\hat{\Gamma}/dz$ of the Higgs boson into a quark or gluon are not given in those references. We calculate the differential decay widths for $H\to Q+X$ and $H \to g+X$ up to order $\alpha_s$ in this work. In the calculation, we neglect the quark mass in the amplitudes and phase space integrals except the quark mass in the Yukawa coupling. This approximation will only lead to an error of ${\cal O}(m_Q^2/m_H^2)$. Then, we have
\begin{widetext}
\begin{eqnarray}
\frac{d\hat{\Gamma}^{\rm NLO}_{H \to Q+X}(y,\mu_F)}{dy}=&&\frac{\sqrt{2} N_c G_F m_H \overline{m}_Q^2(\mu_R)}{8\pi}\Bigg\{\delta(1-y)+\frac{\alpha_s(\mu_R)}{2\pi}\Bigg[P^{(0)}_{QQ}(y)\,{\rm ln}\left(\frac{m_H^2}{\mu_F^2}\right)-3\,C_F \, \delta(1-y) \nonumber \\
&&\times {\rm ln}\left(\frac{m_H^2}{\mu_R^2}\right)+C_Q(y)\Bigg] \Bigg\}, \label{eq.coefun1}\\
\frac{d\hat{\Gamma}^{\rm NLO}_{H \to g+X}(y,\mu_F)}{dy}=&&\sum_{Q=b,c} \frac{\sqrt{2}N_c G_F m_H \overline{m}_Q^2(\mu_R)\alpha_s(\mu_R)}{8\pi^2}\Bigg[P^{(0)}_{gQ}(y)\,{\rm ln}\left(\frac{m_H^2}{\mu_F^2}\right) +C_g(y)\Bigg],\label{eq.coefun2}
\end{eqnarray}
\end{widetext}
where $Q$ can be a quark or an antiquark, $N_c=3$ is the number of quark colors, $C_F=(N_c^2-1)/(2N_c)$ is the quadratic Casimir operator, $G_F$ is the Fermi constant, and $\overline{m}_Q(\mu_R)$ is the running quark mass defined in the modified-minimal-subtraction scheme ($\overline{\rm MS}$). The expressions of the LO splitting functions are
\begin{eqnarray}
P^{(0)}_{QQ}(y) &=& C_F\left[\frac{3}{2}\delta(1-y)+\frac{1+y^2}{(1-y)_+} \right],  \label{eq.spQQ}\\
P^{(0)}_{gQ}(y) &=& C_F \frac{1+(1-y)^2}{y}.
\label{eq.spgQ}
\end{eqnarray}
The expressions of $C_i(y)$ functions in Eqs.(\ref{eq.coefun1}) and (\ref{eq.coefun2}) are
\begin{eqnarray}
C_Q(y)=&& C_F \Bigg\{\left(\frac{3}{2}+\frac{2\pi^2}{3}\right)\delta(1-y)-\frac{3}{2}\frac{1}{(1-y)_+} \nonumber \\
&& +2\left[\frac{{\rm ln}(1-y)}{1-y}\right]_+ +\frac{5}{2}+\frac{y}{2}+4\frac{{\rm ln}\,y}{1-y}\nonumber \\
&& -(1+y)\left[2{\rm ln}\, y+{\rm ln}(1-y)\right] \Bigg\},\end{eqnarray}
\begin{eqnarray}
C_g(y)= C_F \Bigg\{ \frac{1+(1-y)^2}{y}\left[ 2{\rm ln}\, y+ {\rm ln}(1-y) \right]+y \Bigg\}.
\end{eqnarray}
In the calculation, the ultraviolet (UV) divergences are removed by renormalization, and the renormalization of the quark mass is carried out in the usual $\overline{\rm MS}$ scheme. Besides the UV divergences, there are infrared (IR) soft and collinear divergences appearing in the virtual and real corrections. The IR soft divergences are canceled between the virtual and the real corrections, while the IR collinear divergences (which should be absorbed into the bare fragmentation functions) are subtracted according to the $\overline{\rm MS}$ scheme. To avoid large logarithms appearing in $d\hat{\Gamma}_{H \to i+X}/dz$, we will set the renormalization and factorization scales as $\mu_R=\mu_F=m_H$ in the following calculation.

The fragmentation functions for a parton into the $B_c$ meson can be calculated based on the nonrelativistic QCD (NRQCD) factorization theory \cite{Bodwin:1994jh}, i.e.,
\begin{eqnarray}
D_{i\to B_c}(z,\mu_F)=\sum_n d_{i\to (c\bar{b})[n]}(z,\mu_F)\langle{\cal O}^{B_c}(n)\rangle,
\end{eqnarray}
where $d_{i\to c\bar{b}[n]}(z,\mu_F)$ is the short-distance coefficient (SDC) for the $(c\bar{b})[n]$ pair production, which can be calculated through perturbative QCD. $\langle{\cal O}^{B_c}(n)\rangle$ is the long-distance matrix element (LDME) for the transition of a $(c\bar{b})[n]$ pair into the $B_c$ meson, which can be estimated through phenomenological models, e.g. the potential models. The sum extends over intermediate Fock states. In the lowest nonrelativistic approximation, only the Fock state $n=\,^1S^{[1]}_0(^3S^{[1]}_1)$ need to be considered in the production of the $B_c(B_c^*)$ meson.

The LO fragmentation functions for $\bar{b} \to B_c(B_c^*)$ and $c \to B_c(B_c^*)$ were first correctly calculated by the authors of Refs.\cite{Chang:1992bb, Chang:1991bp}. They extracted the fragmentation functions from the processes $Z \to B_c(B_c^*)+b+\bar{c}$ by taking the approximation of $m_{B_c}/m_Z \to 0$. Their results were confirmed by the subsequent calculations in Refs.\cite{Braaten:1993jn, Ma:1994zt} using different methods. For a long time, the NLO fragmentation functions for the $B_c$ production are absent. Recently, with the development of loop-diagram calculation techniques, the NLO fragmentation functions for $\bar{b} \to B_c(B_c^*)$ and $c \to B_c(B_c^*)$ has been given in Ref.\cite{Zheng:2019gnb}. Furthermore, the fragmentation functions for $g\to B_c(B_c^*)$, which start at order $\alpha_s^3$, have been obtained in Refs.\cite{Zheng:2021sdo,Feng:2021qjm}. In this work, we will adopt the fragmentation functions up to order $\alpha_s^3$ obtained in Refs.\cite{Zheng:2019gnb,Zheng:2021sdo}.

In order to avoid large logarithms appearing in $d\hat{\Gamma}_{H \to i+X}/dz$, we have set the factorization scale as $\mu_F=m_H$ in Eq.(\ref{pQCDfact}). However, the large logarithms of $m_H^2/m_{Bc}^2$ will appear in the fragmentation functions. To resum the large logarithms in the fragmentation functions, we first calculate the fragmentation functions up to order $\alpha_s^3$ with initial scales $\mu_{R0}=\mu_{F0}=m_b+m_c$ using the codes developed in our previous works \cite{Zheng:2019gnb,Zheng:2021sdo}. Then the fragmentation functions with $\mu_F=m_H$ can be obtained through solving the DGLAP equations, i.e.,
\begin{eqnarray}
&&\frac{d}{d~{\rm ln}{\mu^2_F}}D_{i \to B_c}(z,\mu_F)\nonumber \\
&&=\frac{\alpha_s(\mu_F)}{2\pi}\sum_j \int_z^1 \frac{dy}{y}P_{ji}(y,\alpha_s(\mu_F)) D_{j \to B_c}(z/y,\mu_F),
\label{eq.dglap}
\end{eqnarray}
where $P_{ji}(y,\alpha_s(\mu_F))$ are the splitting functions, which can be expanded in powers of $\alpha_s$:
\begin{eqnarray}
P_{ji}(y,\alpha_s(\mu_F))=P^{(0)}_{ji}(y)+\frac{\alpha_s(\mu_F)}{2\pi}P^{(1)}_{ji}(y)+{\cal O}(\alpha_s^3).\nonumber \\
\end{eqnarray}
The LO splitting functions for $Q \to Q$ and $Q\to g$ have been given in Eqs.(\ref{eq.spQQ}) and (\ref{eq.spgQ}), and the LO splitting functions for $g \to Q$ and $g \to g$ are
\begin{eqnarray}
P^{(0)}_{Qg}(y) =&& T_F\left[y^2 + (1-y)^2 \right],  \label{eq.spQg}\\
P^{(0)}_{gg}(y) =&& 2C_A \left[ \frac{y}{(1-y)_+}+\frac{1-y}{y}+y(1-y)\right]\nonumber \\
&&+ \frac{1}{6}\delta(1-y)(11C_A-4n_f T_F).
\label{eq.spgg}
\end{eqnarray}
where $C_A=N_c$ and $T_F=1/2$. The NLO corrections to these splitting functions have been obtained in Refs.\cite{Curci:1980uw,Furmanski:1980cm,Floratos:1978ny,Gonzalez-Arroyo:1979qht,Floratos:1981hs}, they are too lengthy to be replicated here.

It is nontrivial to solve these integro-differential equations. In this work, we adopt the program FFEVOL \cite{Hirai:2011si} to solve the DGLAP equations numerically. In solving the DGLAP equations, the NLO fragmentation functions at $\mu_{F0}=m_b+m_c$ are used as the boundary conditions and the NLO spiliting functions are used as the evolution kernel. After the evolution of the fragmentation functions from the initial factorization scale $\mu_{F0}=m_b+m_c$ to the final factorization scale $\mu_F=m_H$, the large logarithms of $m_H^2/(m_b+m_c)^2$ are resummed up to NLL accuracy.

\section{Numerical results and discussion}
\label{sec3}

To do the numerical calculation, the input parameters are adopted as follows:
\begin{eqnarray}
&& G_F=1.16638\times10^{-5}\,{\rm GeV}^{-2}, m_H=125.3\,{\rm GeV}, \nonumber \\
&& \overline{m}_b(\overline{m}_b)=4.18\,{\rm GeV},~~~\overline{m}_c(\overline{m}_c)=1.27\,{\rm GeV},\nonumber \\
&& \vert R_{1S}(0) \vert^2=1.642 ~{\rm GeV^3}, \vert R_{2S}(0) \vert^2=0.983 ~{\rm GeV^3},
\end{eqnarray}
where the values for the Fermi constant and the masses are taken from the Particle Data Group (PDG) \cite{ParticleDataGroup:2022pth}. $R_{1S}(0)$ and $R_{2S}(0)$ are the radial wave functions at the origin for the $(c\bar{b})$ bound states, which are taken from the calculation based on the ${\rm Buchm\ddot{u}ller}$-Tye potential model \cite{Eichten:1995ch}. The running masses at $\mu_R=m_H$ can be obtained through solving the renormalization-group equation, i.e.,
\begin{eqnarray}
\frac{d\, \overline{m}_Q(\mu_R)}{d\,{\rm ln}\mu_R^2}=-\overline{m}_Q(\mu_R)\sum_{i\geq 0}\gamma_{m,i}\left( \frac{\alpha_s(\mu_R)}{\pi}\right)^{i+1},\label{eq.RG}
\end{eqnarray}
where the first two coefficients\cite{Chetyrkin:1997dh,Vermaseren:1997fq} are
\begin{eqnarray}
\gamma_{m,0}&=& 1, \nonumber\\
\gamma_{m,1}&=& \frac{1}{16}\left(\frac{202}{3}-\frac{20}{9}n_f \right),
\end{eqnarray}
and $n_f$ is the number of active flavors. We solve this renormalization-group equation by using the Mathematica package RunDec \cite{Herren:2017osy}, and only the first two coefficients of the right hand of Eq.(\ref{eq.RG}) are preserved (i.e., the obtained running masses at $\mu_R=m_H$ reach the NLL accuracy). Then we have
\begin{eqnarray}
\overline{m}_b(m_H)=2.78\,{\rm GeV},~~~\overline{m}_c(m_H)=0.60\,{\rm GeV}.
\end{eqnarray}
In the calculation of the fragmentation functions in Ref.\cite{Zheng:2019gnb}, the heavy quark masses are renormalized in the on-shell (OS) scheme. The OS (pole) masses for the heavy quarks can be obtained from the $\overline{\rm MS}$ masses through $m_Q=\overline{m}_Q(\overline{m}_Q)[1+4\alpha_s(\overline{m}_Q)/(3\pi)]$ \cite{Gray:1990yh,Broadhurst:1991fy,Chetyrkin:1999ys,Melnikov:2000qh,Marquard:2015qpa}, and we have
\begin{eqnarray}
m_b=4.58\,{\rm GeV},~~~m_c=1.50\,{\rm GeV}.
\end{eqnarray}
For the strong coupling constant, we adopt the two-loop formula, i.e.,
\begin{eqnarray}
\alpha_s(\mu_R)=\frac{4\pi}{\beta_0 L}\left( 1-\frac{\beta_1 {\rm ln}\,L}{\beta_0^2 L}\right),
\end{eqnarray}
where $L={\rm ln}(\mu_R^2/\Lambda_{QCD})$, $\beta_0=11-2n_f/3$, and $\beta_1=102-38n_f/3$. According to $\alpha_s(m_{_Z})=0.1185$, we obtain $\alpha_s(\overline{m}_c)=0.420$, $\alpha_s(\overline{m}_b)=0.228$, $\alpha_s(m_b+m_c)=0.204$, and $\alpha_s(m_H)=0.113$.

\subsection{Comparison of the results at the LO level}

\begin{figure}[htbp]
\centering
\includegraphics[width=0.4\textwidth]{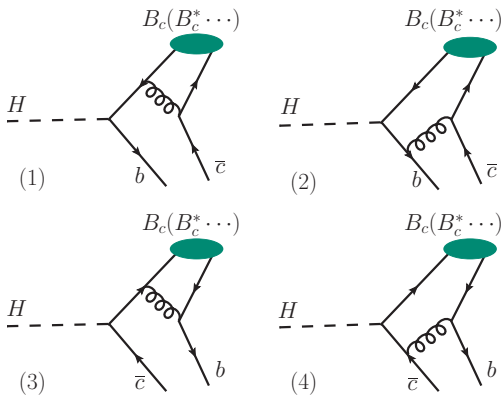}
\caption{LO Feynman diagrams for the decays $H \to B_c(B_c^* \cdots)+X$.
 } \label{feyn}
\end{figure}

In the fragmentation-function approach, some terms which are suppressed by powers of $m_{B_c}^2/m_H^2$ are neglected. In order to see the magnitude of those neglected higher-power (in $m_{B_c}^2/m_H^2$) contributions, we compare the decay widths calculated by the ``direct" NRQCD and the fragmentation-function approaches. Here, we only present the comparison at the LO level.

The differential decay width for the decay $H \to B_c+X$ under the (direct) NRQCD approach \cite{Bodwin:1994jh} can be written as
\begin{eqnarray}
d\Gamma_{H \to B_c+X}=\sum_n d\hat{\Gamma}_{H\to (c\bar{b})[n]+X}\langle{\cal O}^{B_c}(n)\rangle,
\end{eqnarray}
where $d\hat{\Gamma}_{H\to (c\bar{b})[n]+X}$ is the SDC for the $(c\bar{b})[n]$ pair production, which can be calculated through perturbation theory. At the LO level, there are four Feynman diagrams responsible for the decay $H \to B_c+X$. The details about the LO calculation based on the direct NRQCD approach can be found in Ref.\cite{Jiang:2015pah}. In the calculation, we adopt the package FeynArts \cite{Hahn:2000kx} to generate the Feynman diagrams and the amplitudes, and use the FeynCalc \cite{Mertig:1990an,Shtabovenko:2016sxi} to carry out the Dirac traces.

\begin{table}[htb]
\centering{
\begin{tabular}{c c c}
\hline\hline
Contributions &  Direct NRQCD & FF approach \\
\hline
$\overline{b}$-fragmentation & 1.20  &   1.22  \\
$c$-fragmentation & $4.13 \times 10^{-3}$ &  $4.26 \times 10^{-3}$  \\
Interference & $1.25 \times 10^{-2}$ &    \\
Total & 1.22 &  1.22   \\
\hline\hline
\end{tabular}
\caption{The LO partial decay width (unit: keV) for $H \to B_c+X$ under the direct NRQCD approach and the fragmentation-function (FF) approach.}\label{tb.Hbclo}
}
\end{table}

\begin{table}[htb]
\centering{
\begin{tabular}{c c c}
\hline\hline
Contributions &  Direct NRQCD & FF approach \\
\hline
$\overline{b}$-fragmentation & 1.62  &   1.68  \\
$c$-fragmentation & $3.46 \times 10^{-3}$ &  $3.70 \times 10^{-3}$  \\
Interference & $-6.98 \times 10^{-3}$ &    \\
Total & 1.62 &  1.68   \\
\hline\hline
\end{tabular}
\caption{The LO partial decay width (unit: keV) for $H \to B_c^*+X$ under the direct NRQCD approach and the fragmentation-function (FF) approach.}\label{tb.Hbc*lo}
}
\end{table}

The partial decay widths for $H \to B_c+X$ and $H \to B_c^*+X$ under the direct NRQCD approach and the fragmentation-function approach\footnote{It is noted that the calculation under the fragmentation-function approach is restricted to LO here, i.e., the LO fragmentation functions without the DGLAP evolution are directly used in the factorization formalism.} are presented in Tables \ref{tb.Hbclo} and \ref{tb.Hbc*lo}. Here, for consistency, the quark masses are taken as the corresponding pole masses and the strong coupling is taken as $\alpha_s(m_b+m_c)=0.204$ under the two approaches. In the tables, the contributions from the $\bar{b}$-fragmentation and the $c$-fragmentation as well as the total contribution are presented explicitly. For the direct NRQCD approach, the first two Feynman diagrams in Fig.\ref{feyn} are responsible for the $\bar{b}$-fragmentation contribution, while the last two Feynman diagrams are responsible for the $c$-fragmentation contribution\footnote{Any one of the Feynman diagrams in Fig.\ref{feyn} is not gauge invariant. However, the first two diagrams and the last two diagrams constitute two gauge-invariant subgroups, respectively. Hence, in an arbitrary covariant gauge (e.g. the Feynman gauge), the first (last) two Feynman diagrams in Fig.\ref{feyn} should be simultaneity taken into account for the $\bar{b}$-fragmentation ($c$-fragmentation) contribution.}. The interference contribution comes from the interference of the first two Feynman diagrams and the last two Feynman diagrams, and the fragmentation-function approach can not give the interference contribution.

From Tables \ref{tb.Hbclo} and \ref{tb.Hbc*lo}, we can see that the differences between the decay widths under two approaches are very small for both the $\bar{b}$-fragmentation and the $c$-fragmentation.\footnote{The results show that the difference between the two approaches for $B_c^*$ is larger than that for $B_c$. This indicates that the accuracy of the fragmentation-function approximation depends not only on the large energy scale involved in the process (e.g. $m_H$ in $H \to B_c(B_c^*)+X$), but also on the quantum number of the produced state. This point is also shown in the process $gg\to B_c(B_c^*)+b+\bar{c}$ \cite{Chang:1996jt}.} Moreover, the interference contributions are very small in both the $B_c$ and $B_c^*$ cases. There are two reasons for the small interference contributions. First, the interference contributions come from the interference of the $\bar{b}$-fragmentation diagrams and the $c$-fragmentation diagrams. The $c$-fragmentation diagrams are suppressed compared to the $\bar{b}$-fragmentation diagrams. This leads to the interference contributions are suppressed compared to the $\bar{b}$-fragmentation contributions. Second, the dominant contribution of the $\bar{b}$-fragmentation diagrams comes from the phase-space region where the relative velocity of the $B_c(B_c^*)$ meson and the $\bar{c}$-quark is small and back-to-back with the $b$-quark, while the dominant contribution of the $c$-fragmentation diagrams comes from the phase-space region where the relative velocity of the $B_c(B_c^*)$ meson and the $b$-quark is small and back-to-back with the $\bar{c}$-quark, i.e., the dominant contributions of the $\bar{b}$-fragmentation diagrams and the $c$-fragmentation diagrams come from different phase-space regions, which further suppresses the contributions of the interference. These two reasons lead to the very small interference contributions. Therefore, the higher-power terms neglected in the fragmentation-function approach are very small in the two decay processes, the fragmentation-function approach can give a good approximation to the direct NRQCD approach.

The results show that the $c$-fragmentation contributions are about 3 orders of magnitude smaller than the corresponding $\bar{b}$-fragmentation contributions. There are two reasons for the very small $c$-fragmentation contributions: One is that the magnitude of the Yukawa coupling of $Hc\bar{c}$ is smaller than that of $Hb\bar{b}$,  and the other is that the fragmentation probability of $c \to B_c$ is smaller than that of $\bar{b} \to B_c$.

\begin{figure}[htbp]
\centering
\includegraphics[width=0.45\textwidth]{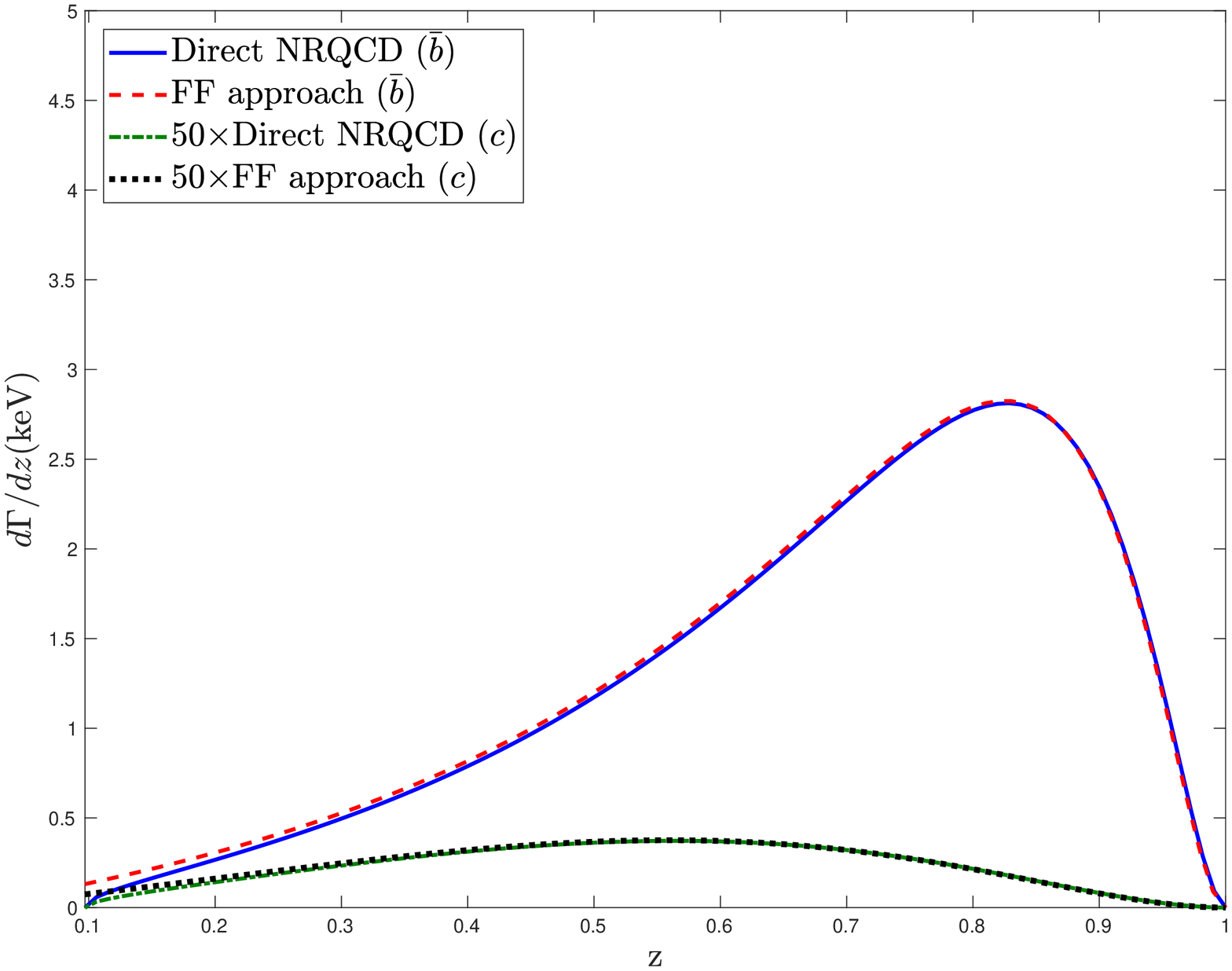}
\includegraphics[width=0.45\textwidth]{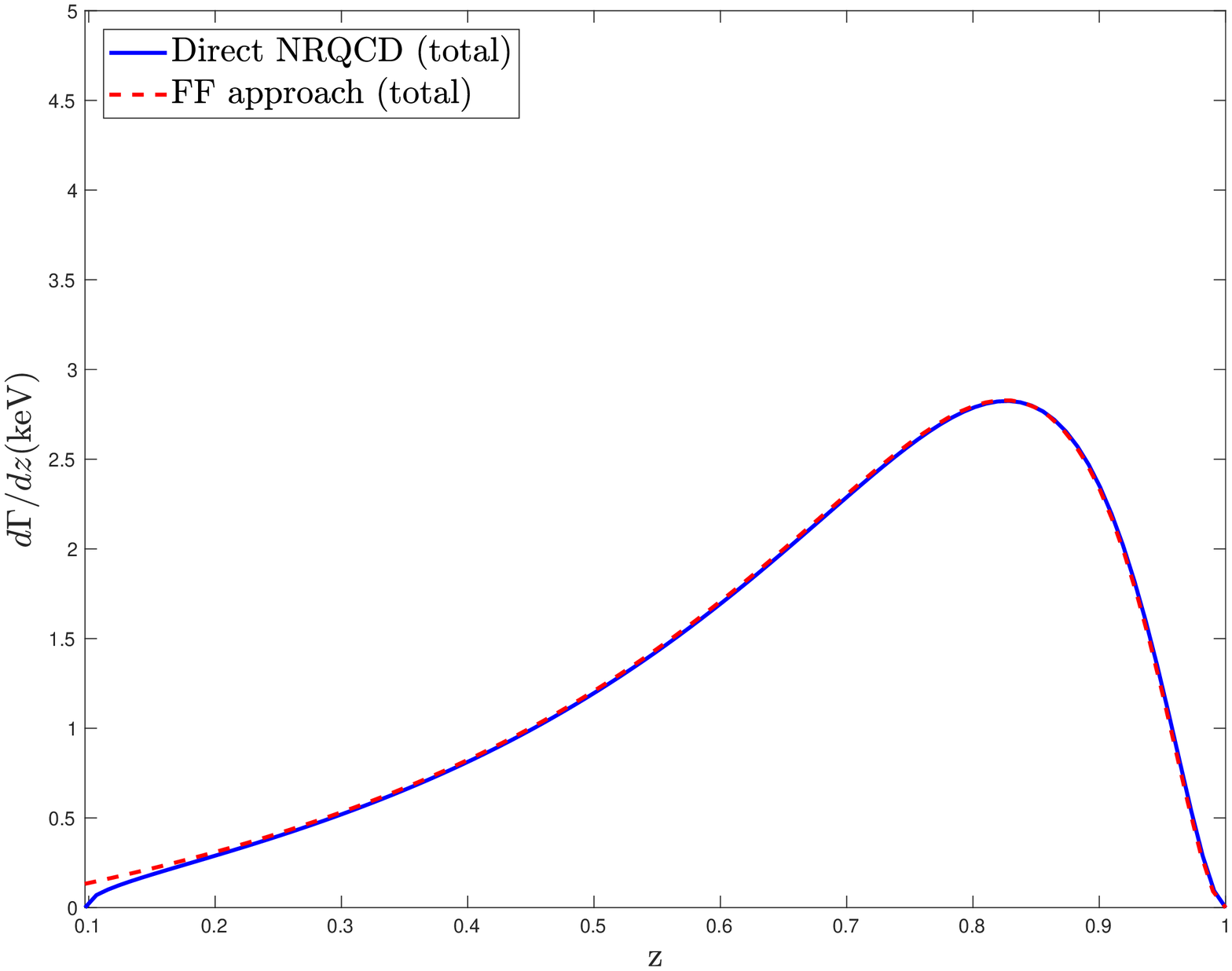}
\caption{Comparison of differential decay widths $d\Gamma/dz$ of $H \to B_c+X$ calculated based on the direct NRQCD approach and the fragmentation-function (FF) approach. The upper one shows the contributions of $\bar{b}$-fragmentation and $c$-fragmentation respectively, the lower one shows total contribution. In order to put the results from the $\bar{b}$-fragmentation and the $c$-fragmentation into one figure, the curves for the $c$-fragmentation are multiplied by a factor of 50.}
\label{gamma1s0lo}
\end{figure}

\begin{figure}[htbp]
\centering
\includegraphics[width=0.45\textwidth]{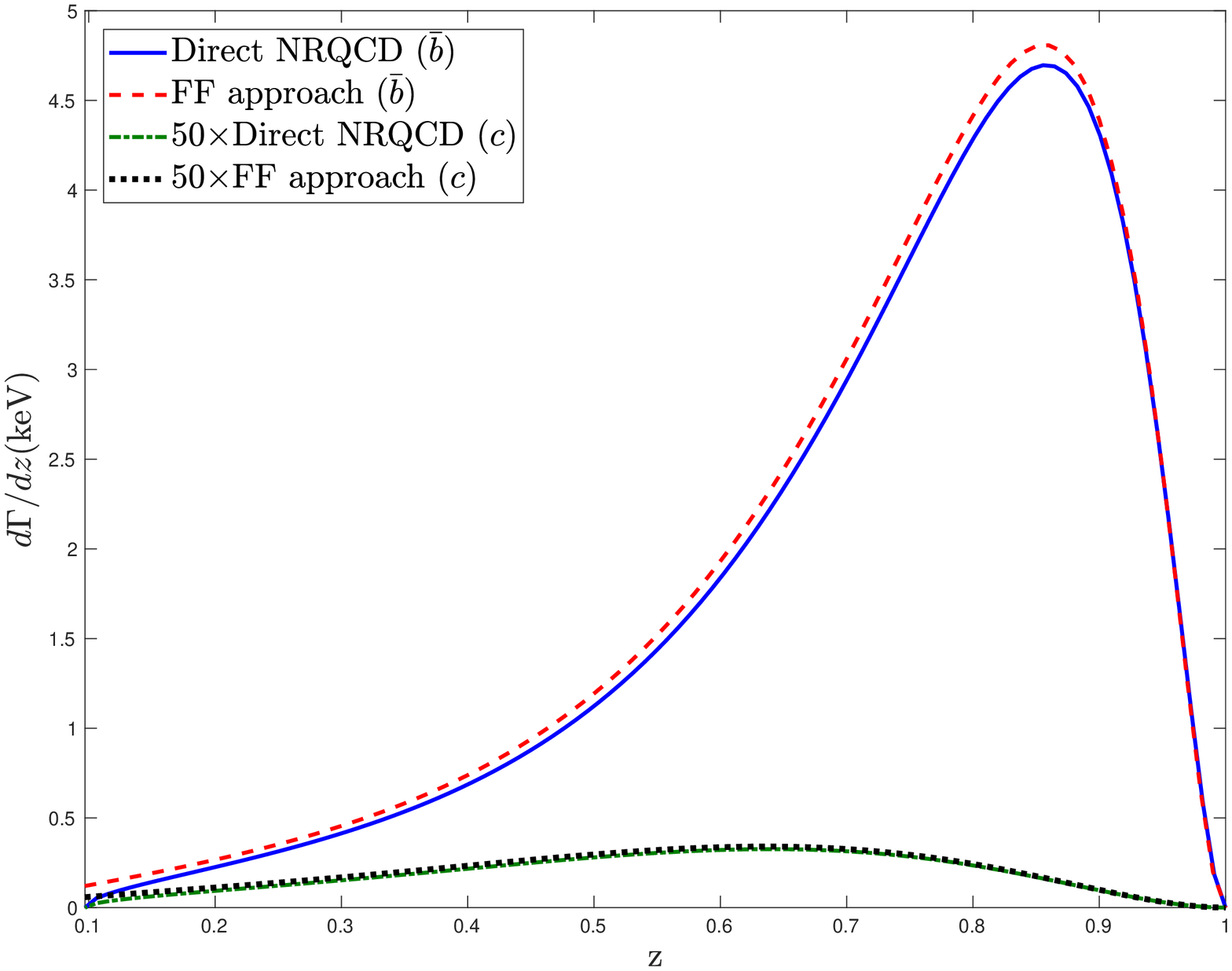}
\includegraphics[width=0.45\textwidth]{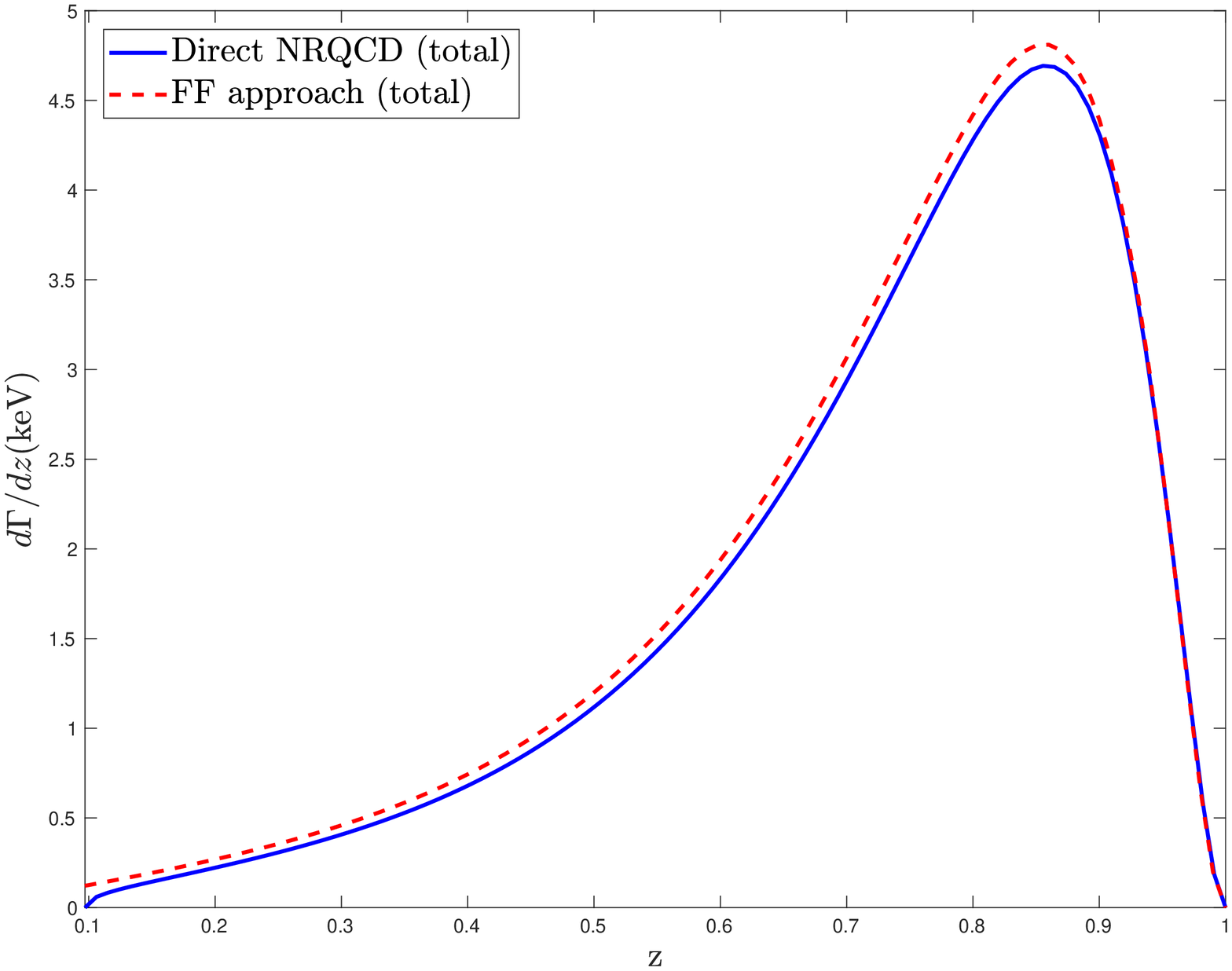}
\caption{Comparison of differential decay widths $d\Gamma/dz$ of $H \to B_c^*+X$ calculated based on the direct NRQCD approach and the fragmentation-function (FF) approach. The upper one shows the contributions of $\bar{b}$-fragmentation and $c$-fragmentation respectively, the lower one shows total contribution. In order to put the results from the $\bar{b}$-fragmentation and the $c$-fragmentation into one figure, the curves for the $c$-fragmentation are multiplied by a factor of 50.}
\label{gamma3s1lo}
\end{figure}

The differential decay widths $d\Gamma/dz$ of $H \to B_c+X$ and $H \to B_c^*+X$ under the direct NRQCD approach and the fragmentation-function approach are shown in Figs. \ref{gamma1s0lo} and \ref{gamma3s1lo}. From the figures, we can see that the curves from the two approaches are very close. The differences between the two approaches are relatively small at large $z$ values, and relatively large at small $z$ values. The reason is that the fragmentation contribution mainly comes from the phase-space regions with large $z$ value. More explicitly, the contribution of $\bar{b}$-fragmentation comes from the phase-space region where the $B_c(B_c^*)$-$\bar{c}$ system has small invariant mass and large momentum. The contribution of $c$-fragmentation comes from the phase-space region where the $B_c(B_c^*)$-$b$ system has small invariant mass and large momentum. In these phase-space regions, the fragmentation-function approach can well describe the process with the $B_c(B_c^*)$ production. In the phase-space regions where the momentum of the $B_c(B_c^*)$ meson is very small, the non-fragmentation contribution becomes relatively important.

\subsection{Contribution from the $Ht\bar{t}$ coupling}

\begin{figure}[htbp]
\centering
\includegraphics[width=0.45\textwidth]{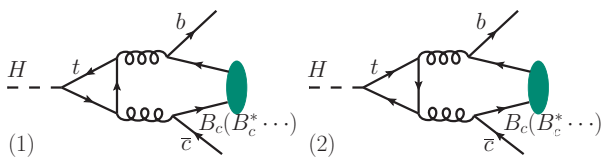}
\caption{Feynman diagrams induced by the triangle top-quark loop.
 } \label{feynHtt}
\end{figure}

From the comparison of the LO results under the two approaches given in the last subsection, we found that the fragmentation-function approach (without the resummation of large logarithms) can give a good approximation to the direct NRQCD approach, i.e., the power corrections in the fragmentation-function approach are negligible at the LO level. At the NLO, there are nonfragmentation Feynman diagrams induced by a triangle top-quark loop as shown in Fig.\ref{feynHtt}. Compared with the fragmentation contribution, the contribution from these Feynman diagrams is suppressed by powers of $m_{B_c}^2/m_H^2$ but enhanced by the $Ht\bar{t}$ coupling. Therefore, before giving the results from the fragmentation-function approach up to the NLL accuracy, it is important to see how much these triangle top-loop diagrams contribute. In fact, the authors of Ref.\cite{Jiang:2015pah} have calculated the contribution from the triangle top-quark loop diagrams. They obtained very strange results, i.e., the triangle top-loop contribution in the $B_c^*$ case is one order of magnitude smaller than that in the $B_c$ case. To further illustrate the reason of the smallness of the triangle top-loop contribution in the $B_c^*$ case, we recalculate the contribution from the triangle top-loop diagrams here. Furthermore, the authors of Ref.\cite{Jiang:2015pah} found that the interference contribution between Fig.\ref{feynHtt} and Fig.\ref{feyn} is very small for both the $B_c$ and $B_c^*$ cases. However, we can not conclude from the small interference contribution that the contribution from the square of Fig.\ref{feynHtt} is also small. Because the topologies of Fig.\ref{feynHtt} and Fig.\ref{feyn} are significantly different, they may be dominated by different phase-space regions. Hence, in addition to the contribution from the interference between the Fig.\ref{feynHtt} and Fig.\ref{feyn}, we also calculate the contribution from the square of Fig.\ref{feynHtt}.

\begin{table}[htb]
\centering{
\begin{tabular}{c c c}
\hline\hline
Contributions &  $B_c$ & $B_c^*$ \\
\hline
Interference of Fig.\ref{feyn} and Fig.\ref{feynHtt} & $4.39 \times 10^{-2}$  &   $5.43 \times 10^{-3}$  \\
Square of Fig.\ref{feyn} & $2.04 \times 10^{-3}$ &  $5.09 \times 10^{-3}$  \\
\hline\hline
\end{tabular}
\caption{Contributions (unit:keV) of the triangle top-loop diagrams to decay widths.}\label{tb.Hbc*top}
}
\end{table}

\begin{figure}[htbp]
\centering
\includegraphics[width=0.45\textwidth]{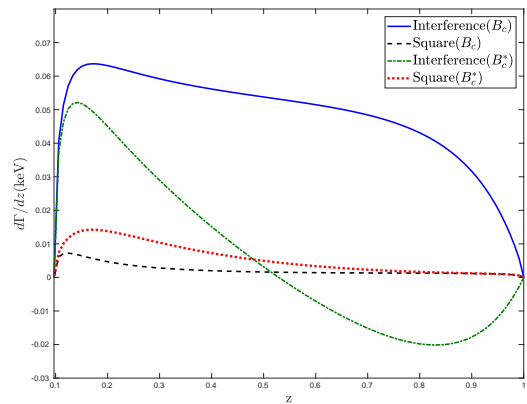}
\caption{Contributions of the triangle top-loop diagrams  to $d\Gamma/dz$, where ``Interference" denotes the contribution coming from the interference between Fig.\ref{feyn} and Fig.\ref{feynHtt}, and ``Square" denotes the contribution coming from the square of Fig.\ref{feynHtt}.
 } \label{gammaztop}
\end{figure}

In Table \ref{tb.Hbc*top} and Fig.\ref{gammaztop}, the contributions from the triangle top-quark loop are presented\footnote{Adopting the same input parameters as in Ref.\cite{Jiang:2015pah}, we are able to reproduce the numerical results for the contribution of the interference between the triangle top-quark loop diagrams and the LO diagrams given in Table IV of Ref.\cite{Jiang:2015pah}.}. In the calculation, the top-quark mass is taken as $m_t=172.8\,{\rm GeV}$ \cite{ParticleDataGroup:2022pth}, and the other parameters are taken the same values as those in the last subsection. From Table \ref{tb.Hbc*top}, we can see that the interference contribution in the $B_c^*$ case is one order of magnitude smaller than that in the $B_c$ case. This can be understood by the distribution shown in Fig.\ref{gammaztop}. The distribution of the interference contribution in the $B_c^*$ case is negative at large $z$ values, this indicates that there is a large cancellation between the contributions from different phase-space regions. Furthermore, we can also see that the contributions (the interference contribution as well as the squared contribution) from the triangle top-quark loop are very small compared with the LO contributions.

\subsection{Results up to NLL accuracy under the fragmentation-function approach}

\begin{widetext}

\begin{table}[htb]
\centering{
\begin{tabular}{c c c c c}
\hline\hline
Contributions &  $B_c$~ & ~$B_c^*$~&  $B_c(2^1S_0)$~ & ~$B_c^*(2^3S_1)$~ \\
\hline
$\overline{b}$-fragmentation & 0.673  &   0.766  & 0.403  &   0.459 \\
$c$-fragmentation & $1.47 \times 10^{-3}$ &  $1.25 \times 10^{-3}$ & $8.80 \times 10^{-4}$ &  $7.48 \times 10^{-4}$  \\
$g$-fragmentation & $-1.80 \times 10^{-3}$ &  $-2.45 \times 10^{-3}$ & $-1.07 \times 10^{-3}$ &  $-1.47 \times 10^{-3}$  \\
Triangle top-loop & $4.59 \times 10^{-2}$ &  $1.05 \times 10^{-2}$ & $ 2.75\times 10^{-2}$ &  $ 6.29\times 10^{-3}$  \\
Total & 0.719 &  0.775 & 0.430 & 0.465  \\
\hline\hline
\end{tabular}
\caption{The partial decay widths (unit: keV) for the Higgs decays to $B_c$, $B_c^*$, $B_c(2^1S_0)$ and $B_c^*(2^3S_1)$, where the contributions from three fragmentation channels and the triangle top-loop contribution are given explicitly.}\label{tb.HbcNLL}
}
\end{table}

\end{widetext}

\begin{figure}[htbp]
\centering
\includegraphics[width=0.45\textwidth]{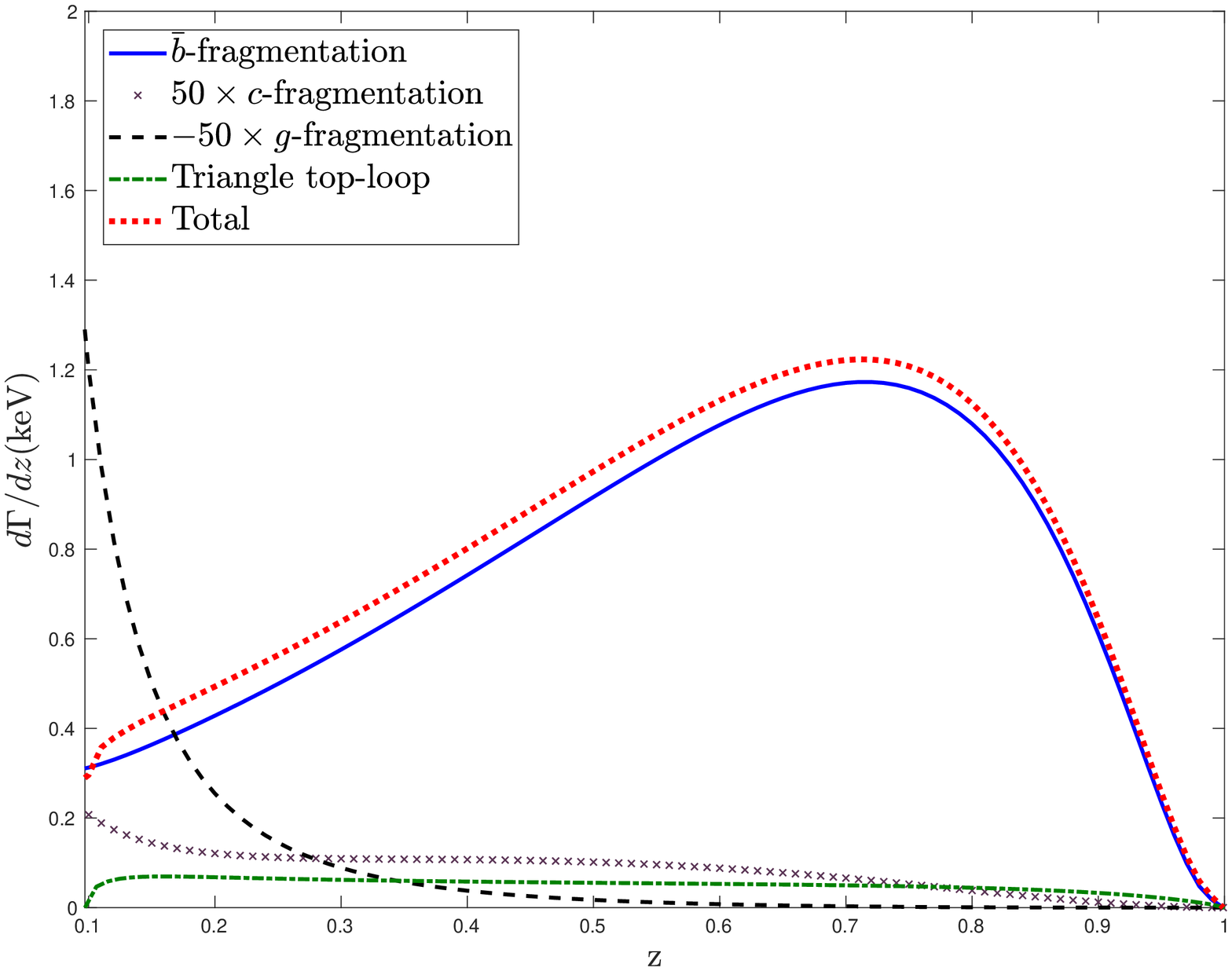}
\caption{The differential decay width $d\Gamma/dz$ for $H\to B_c+X$, where the contributions from the three fragmentation channels and the triangle top-quark loop are shown explicitly. In order to put the results into one figure, the curve for the $c$-fragmentation is multiplied by a factor of 50, and the curve for the $g$-fragmentation is multiplied by a factor of -50.} \label{gammaz}
\end{figure}

\begin{figure}[htbp]
\centering
\includegraphics[width=0.45\textwidth]{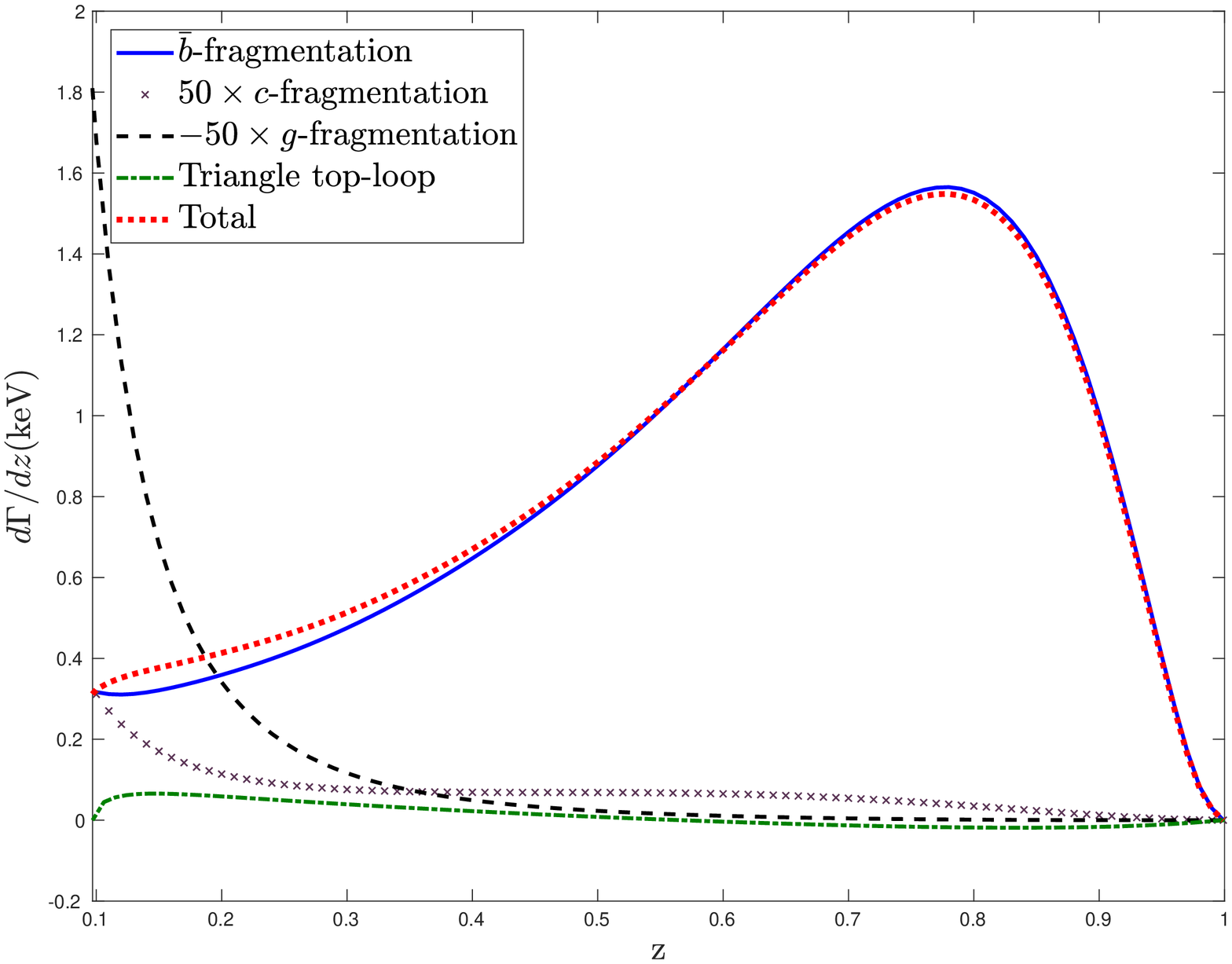}
\caption{The differential decay width $d\Gamma/dz$ for $H\to B_c^*+X$, where the contributions from the three fragmentation channels and the triangle top-quark loop are shown explicitly. In order to put the results into one figure, the curve for the $c$-fragmentation is multiplied by a factor of 50, and the curve for the $g$-fragmentation is multiplied by a factor of -50.} \label{gammaz3s1}
\end{figure}

\begin{figure}[htbp]
\centering
\includegraphics[width=0.45\textwidth]{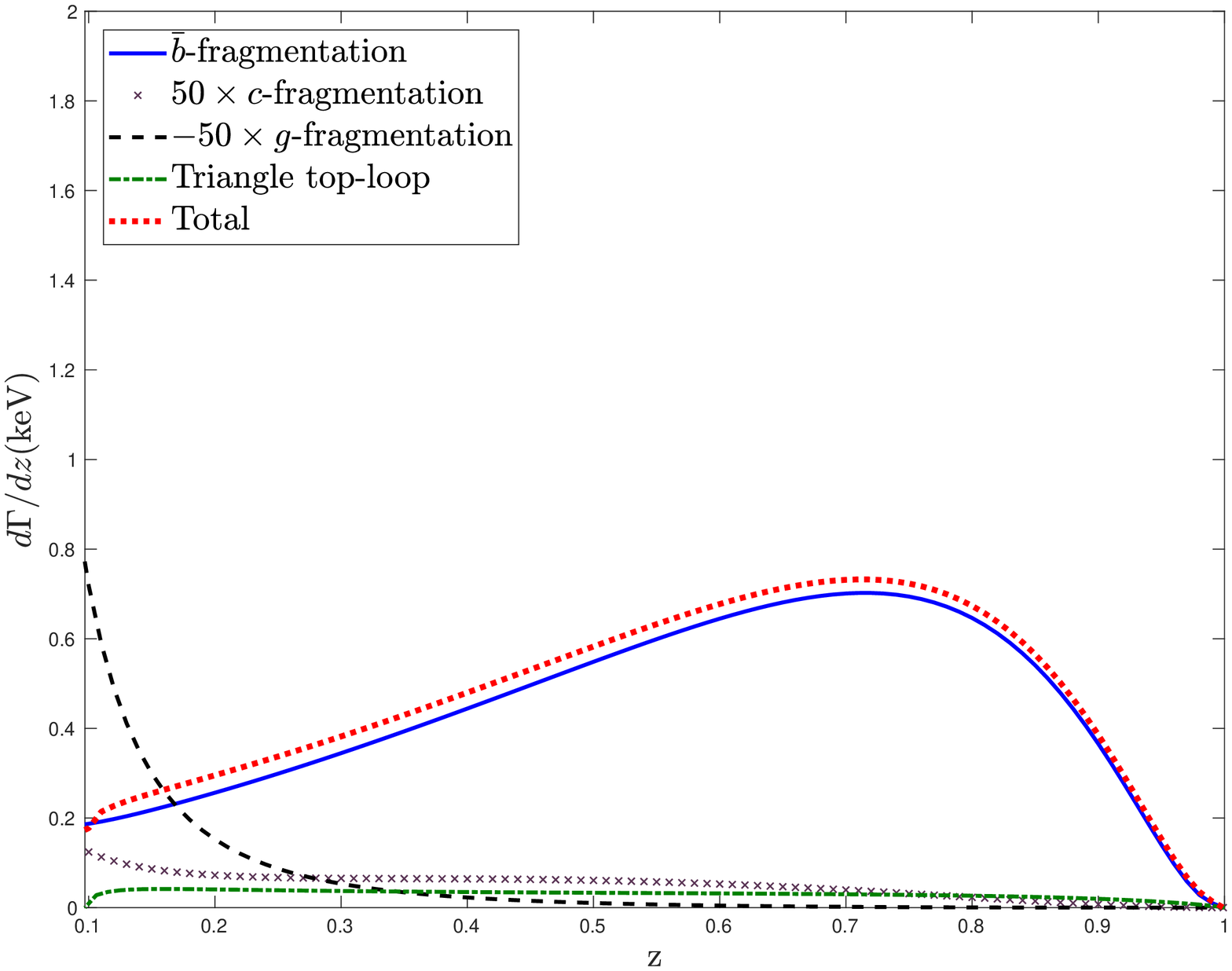}
\caption{The differential decay width $d\Gamma/dz$ for $H\to B_c(2^1S_0)+X$, where the contributions from the three fragmentation channels and the triangle top-quark loop are shown explicitly. In order to put the results into one figure, the curve for the $c$-fragmentation is multiplied by a factor of 50, and the curve for the $g$-fragmentation is multiplied by a factor of -50.} \label{gammaz-2s}
\end{figure}

\begin{figure}[htbp]
\centering
\includegraphics[width=0.45\textwidth]{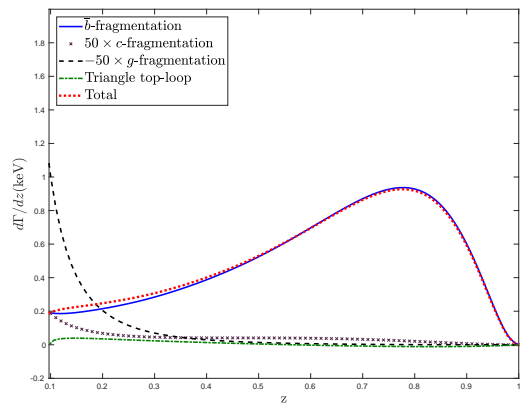}
\caption{The differential decay width $d\Gamma/dz$ for $H\to B_c^*(2^3S_1)+X$, where the contributions from the three fragmentation channels and the triangle top-quark loop are shown explicitly. In order to put the results into one figure, the curve for the $c$-fragmentation is multiplied by a factor of 50, and the curve for the $g$-fragmentation is multiplied by a factor of -50.} \label{gammaz3s1-2s}
\end{figure}

From the analysis presented in the above two subsections, we believe that the neglected power suppressed terms in the fragmentation-function approach are small in the decays $H\to B_c+X$ and $H\to B_c^*+X$ even up to the NLO level. In this subsection, we present the decay widths calculated based on the the fragmentation-function approach up to the NLL accuracy.

In Table \ref{tb.HbcNLL}, the partial decay widths for the Higgs decays to $B_c$, $B_c^*$, $B_c(2^1S_0)$ and $B_c^*(2^3S_1)$ are presented, where the contributions from the different fragmentation channels and the contribution from the triangle top-quark loop are given explicitly. The calculation method for these fragmentation contributions have been described in Sec.\ref{sec2}, i.e., the factorization scale is taken as $\mu_F=m_H$ in Eq.(\ref{pQCDfact}) and the fragmentation functions at $\mu_F=m_H$ are obtained through the DGLAP evolution from the initial factorization scale $\mu_{F0}=m_b+m_c$. Hence, the large logarithms of $m_H^2/(m_b+m_c)^2$ which arise from the collinear gluon emission and the renormalization of the Yukawa couplings have been resummed in the numerical results for these fragmentation contributions. The input parameters have been given at the beginning of this section.

From Table \ref{tb.HbcNLL}, we can see that the decay widths up to NLL accuracy are significantly smaller than the corresponding LO decay widths shown in Tables \ref{tb.Hbclo} and \ref{tb.Hbc*lo}. This indicates that the higher-order corrections, especially the large logarithmic terms, are important in these decay processes. Therefore, the resummation of large logarithms should be taken into consideration for giving high-precision predictions. The contributions from the $c$-fragmentation and the $g$-fragmentation are very small compared to the $\bar{b}$-fragmentation contribution. Moreover, the $g$-fragmentation contribution is negative for these processes.

In Figs.\ref{gammaz}, \ref{gammaz3s1}, \ref{gammaz-2s} and \ref{gammaz3s1-2s}, the differential decay widths $d\Gamma/dz$ for the $B_c$, $B_c^*$, $B_c(2^1S_0)$ and $B_c^*(2^3S_1)$ states are shown. In the figures, the contributions of the different fragmentation channels and the triangle top-quark loop to $d\Gamma/dz$ are shown explicitly. From these figures, we can see that the differential decay widths are dominated by the $\bar{b}$-fragmentation contribution for all the $z$ values. The $c$-fragmentation and the $g$-fragmentation contributions mainly come from the small $z$ values. However, even for small $z$ values, these contributions are also small compared to the $\bar{b}$-fragmentation contribution.

\subsection{Uncertainty analysis}

In this subsection, we will estimate the theoretical uncertainties for these partial decay widths. The main uncertainty sources for these decay widths include the factorization and renormalization scales, the heavy quark masses, the Higgs boson mass, and the $c\bar{b}$ radial wave functions at the origin. The dependence of the decay widths on the Higgs boson mass mainly comes from the partonic decay widths $d \hat{\Gamma}_{H \to i+X}(y,\mu_F)/dy(i=Q,g)$, which contain a global factor $m_H$. The uncertainty for the world average value of the Higgs boson mass given by the PDG is about $0.2\,{\rm GeV}$ \cite{ParticleDataGroup:2022pth}, which is only about $0.2\%$ of the Higgs boson mass. Therefore, the uncertainties for the decay widths caused by the Higgs boson mass are only about $0.2\%$ of their central values. Since the uncertainties caused by the Higgs mass are very small, we will neglect this uncertainty source in the following uncertainty estimaton.

There are several factorization and renormalization scales involved in the calculation based on the fragmentation-function approach: the initial (lower) factorization and renormalization scales ($\mu_{F0}$ and $\mu_{R0}$) for the initial fragmentation functions; the final (upper) factorization and renormalization scales ($\mu_{F}$ and $\mu_{R}$). In the calculation presented in the last subsection, these scales are set as $\mu_{F0}=\mu_{R0}=m_b+m_c$ and $\mu_{F}=\mu_{R}=m_H$. In the uncertainty estimation, we vary them by a factor 2 from their central values, i.e., $\mu_{F0}=\mu_{R0}\in [(m_b+m_c)/2,2(m_b+m_c)]$ and $\mu_{F}=\mu_{R}\in [m_H/2,2m_H]$. Then we obtain the uncertainties caused by the lower and upper scales:
\begin{eqnarray}
&&\Gamma_{H\to B_c+X}=0.719^{+0.087+0.037}_{-0.099-0.032}\,{\rm keV},\nonumber\\
&&\Gamma_{H\to B^*_c+X}=0.775^{+0.008+0.041}_{-0.072-0.035}\,{\rm keV},\nonumber\\
&&\Gamma_{H\to B_c(2\,^{1}S_0)+X}=0.430^{+0.053+0.022}_{-0.059-0.019}\,{\rm keV},\nonumber\\
&&\Gamma_{H\to B^*_c(2\,^{3}S_1)+X}=0.465^{+0.004+0.024}_{-0.044-0.022}\,{\rm keV},
\label{uncert.scale}
\end{eqnarray}
where the first uncertainty is caused by the lower factorization and renormalization scales, and the second uncertainty is caused by the upper factorization and renormalization scales.

For the uncertainties caused by the heavy quark masses, we estimate them by using the errors $\overline{m}_b(\overline{m}_b)=4.18^{+0.04}_{-0.03}{\rm GeV}$ and $\overline{m}_c(\overline{m}_c)=1.27\pm 0.02{\rm GeV}$ given by the PDG. We obtain the uncertainties caused by the heavy quark masses:
\begin{eqnarray}
&&\Gamma_{H\to B_c+X}=0.719^{+0.007+0.029}_{-0.005-0.035}\,{\rm keV},\nonumber\\
&&\Gamma_{H\to B^*_c+X}=0.775^{+0.012+0.037}_{-0.007-0.042}\,{\rm keV},\nonumber\\
&&\Gamma_{H\to B_c(2\,^{1}S_0)+X}=0.430^{+0.005+0.018}_{-0.003-0.021}\,{\rm keV},\nonumber\\
&&\Gamma_{H\to B^*_c(2\,^{3}S_1)+X}=0.465^{+0.006+0.021}_{-0.005-0.026}\,{\rm keV},
\label{uncert.mass}
\end{eqnarray}
where the first uncertainty is caused by the bottom quark mass, while the second uncertainty is caused by the charm quark mass.

\begin{table}[htb]
\centering{
\begin{tabular}{c c c c}
\hline\hline
Level &  BT (\cite{Eichten:1995ch}) & Logarithmic (\cite{Eichten:1995ch}) & Cornell (\cite{Eichten:2019gig}) \\
\hline
1S & 1.642  & 1.508  & 1.994 \\
2S & 0.983  & 0.770  & 1.144 \\
\hline\hline
\end{tabular}
\caption{Model dependence of the radial wave functions at the origin (unit: ${\rm GeV}^3$) for the $c\bar{b}$ mesons, where ``BT" denotes the ${\rm Buchm\ddot{u}ller}$-Tye potential model.}\label{tb.wave-function}
}
\end{table}

For the wave functions at the origin, we have adopted the values based on the ${\rm Buchm\ddot{u}ller}$-Tye potential model given in Ref.\cite{Eichten:1995ch}. However, the authors of Ref.\cite{Eichten:1995ch} did not give an error estimate to the wave functions. In order to give an estimate to the uncertainties from the wave functions, we take the values based on the ${\rm Buchm\ddot{u}ller}$-Tye potential model as the central values, while take the values based on the logarithmic potential and the Cornell potential as the boundary values for the wave functions. The values for the radial wave functions based on the three potential models are shown in Table \ref{tb.wave-function}. We obtain the uncertainties caused by the wave functions as follows\footnote{The wave functions at the origin are overall factors in the calculation. If we have more accurate values of the wave functions (e.g., extracted from the experimental data) in the future, we can easily update our results based on the more accurate values of the wave functions.}:
\begin{eqnarray}
&&\Gamma_{H\to B_c+X}=0.719^{+0.154}_{-0.059}\,{\rm keV},\nonumber\\
&&\Gamma_{H\to B^*_c+X}=0.775^{+0.166}_{-0.063}\,{\rm keV},\nonumber\\
&&\Gamma_{H\to B_c(2\,^{1}S_0)+X}=0.430^{+0.070}_{-0.093}\,{\rm keV},\nonumber\\
&&\Gamma_{H\to B^*_c(2\,^{3}S_1)+X}=0.465^{+0.076}_{-0.101}\,{\rm keV}.
\label{uncert.wavefunc}
\end{eqnarray}

Adding the uncertainties from different sources in quadrature, we obtain the total theoretical uncertainties for these partial decay widths as follows:
\begin{eqnarray}
&&\Gamma_{H\to B_c+X}=0.719^{+0.183}_{-0.125}\,{\rm keV},\nonumber\\
&&\Gamma_{H\to B^*_c+X}=0.775^{+0.176}_{-0.110}\,{\rm keV},\nonumber\\
&&\Gamma_{H\to B_c(2\,^{1}S_0)+X}=0.430^{+0.092}_{-0.114}\,{\rm keV},\nonumber\\
&&\Gamma_{H\to B^*_c(2\,^{3}S_1)+X}=0.465^{+0.083}_{-0.115}\,{\rm keV}.
\label{eq.uncert.total}
\end{eqnarray}

\section{Summary}
\label{sec4}

In the present paper, we have calculated the partial decay widths for the Higgs boson decays to the $B_c$, $B_c^*$, $B_c(2\,^{1}S_{0})$, and $B_c^*(2\,^{3}S_{1})$ mesons based on the fragmentation-function approach. The decay widths and the differential distributions are obtained, and the theoretical uncertainties for the decay widths are estimated. In the calculation, the fragmentation functions up to order $\alpha_s^3$ for the $B_c(B_c^*)$ production calculated in the previous works are used as the initial fragmentation functions. The large logarithms that arise from the renormalization of the Yukawa couplings and the collinear gluon emissions are resummed up to NLL accuracy through solving the evolution equations of the running heavy-quark masses and the fragmentation functions. After including these higher-order contributions, the decay widths are reduced significantly compared to the LO predictions.

In order to have a feel on the size of the higher-power terms (in $m_{B_c}^2/m_H^2$) that neglected in the fragmentation-function approach, we compared the decay widths under the direct NRQCD approach with those under the fragmentation-function approach at the LO level. The results show that the decay widths under the two approaches are very close to each other, i.e., those higher-power terms are very small at the LO level. We also studied the contributions induced by the triangle top-quark loop, which are enhanced by the $Ht\bar{t}$ coupling. The results show that these contributions are very small compared to the fragmentation contributions. Moreover, we found that the interference between the triangle top-quark loop diagrams and the LO Feynman diagrams for $H\to B_c^*+X$ has a strong cancellation between different phase-space regions. This leads to a much smaller contribution from the triangle top-quark loop in the $B_c^*$ case than that in the $B_c$ case.

Since the $B_c$ excited states below the BD threshold will decay to the ground state $B_c$ with almost $100\%$ probability, the total decay width for the Higgs boson decay to the $B_c$ meson is approximately equal to the sum of the decay widths for the Higgs boson decays to the $c\bar{b}$ meson states below the BD threshold. Adding the decay widths for the $S$-wave states ($B_c$, $B_c^*$, $B_c(2\,^{1}S_{0})$, and $B_c^*(2\,^{3}S_{1})$) shown in Eq.(\ref{eq.uncert.total}), and using the $\Gamma_{H}\approx 3.2\,{\rm MeV}$ \cite{Eichten:1995ch}, we obtain that the total branching fraction for the Higgs boson decays into the $B_c$ meson is about $7.47\times10^{-4}$. According to the total branching fraction, we estimate that there are about $1.2 \times 10^5$ $B_c$ events will be produced via the Higgs boson decays at the HL-LHC with $3\,ab^{-1}$, and about $1.6 \times 10^6$ $B_c$ events will be produced via the Higgs boson decays at the HE-LHC with $15\,ab^{-1}$. Therefore, these decay processes may be studied at the future HL-LHC and HE-LHC, and provide a complementary method for measuring the bottom-quark Yukawa coupling.

\hspace{2cm}

\noindent {\bf Acknowledgments:} This work was supported in part by the Natural Science Foundation of China under Grants No.12005028, No.12175025 and No.12147102, by the China Postdoctoral Science Foundation under Grant No.2021M693743, by the Fundamental Research Funds for the Central Universities under Grant No.2020CQJQY-Z003, by the Chongqing Natural Science Foundation under Grant No.CSTB2022NSCQ-MSX0415, and by the Chongqing Graduate Research and Innovation Foundation under Grant No.ydstd1912.

\hspace{2cm}


\begin{thebibliography}{1}


\bibitem{ATLAS:2012yve}
G.~Aad \textit{et al.} [ATLAS],
Observation of a new particle in the search for the Standard Model Higgs boson with the ATLAS detector at the LHC,
Phys. Lett. B \textbf{716}, 1-29 (2012).

\bibitem{CMS:2012qbp}
S.~Chatrchyan \textit{et al.} [CMS],
Observation of a New Boson at a Mass of 125 GeV with the CMS Experiment at the LHC,
Phys. Lett. B \textbf{716}, 30-61 (2012).

\bibitem{ATLAS:2015muc}
G.~Aad \textit{et al.} [ATLAS],
Study of (W/Z)H production and Higgs boson couplings using $H \rightarrow WW^{\ast}$ decays with the ATLAS detector,
JHEP \textbf{08}, 137 (2015).

\bibitem{ATLAS:2018pgp}
M.~Aaboud \textit{et al.} [ATLAS],
Combined measurement of differential and total cross sections in the $H \rightarrow \gamma \gamma$ and the $H \rightarrow ZZ^* \rightarrow 4\ell$ decay channels at $\sqrt{s} = 13$ TeV with the ATLAS detector,
Phys. Lett. B \textbf{786}, 114-133 (2018).

\bibitem{CMS:2018zzl}
A.~M.~Sirunyan \textit{et al.} [CMS],
Measurements of properties of the Higgs boson decaying to a W boson pair in pp collisions at $\sqrt{s}=$ 13 TeV,
Phys. Lett. B \textbf{791}, 96 (2019).

\bibitem{ATLAS:2016neq}
G.~Aad \textit{et al.} [ATLAS and CMS],
Measurements of the Higgs boson production and decay rates and constraints on its couplings from a combined ATLAS and CMS analysis of the LHC pp collision data at $ \sqrt{s}=7 $ and 8 TeV,
JHEP \textbf{08}, 045 (2016).

\bibitem{ATLAS:2018mme}
M.~Aaboud \textit{et al.} [ATLAS],
Observation of Higgs boson production in association with a top quark pair at the LHC with the ATLAS detector,
Phys. Lett. B \textbf{784}, 173-191 (2018).

\bibitem{CMS:2018uxb}
A.~M.~Sirunyan \textit{et al.} [CMS],
Observation of $\mathrm{t\overline{t}}$H production,
Phys. Rev. Lett. \textbf{120}, 231801 (2018).

\bibitem{ATLAS:2018kot}
M.~Aaboud \textit{et al.} [ATLAS],
Observation of $H \rightarrow b\bar{b}$ decays and $VH$ production with the ATLAS detector,
Phys. Lett. B \textbf{786}, 59-86 (2018).

\bibitem{CMS:2018nsn}
A.~M.~Sirunyan \textit{et al.} [CMS],
Observation of Higgs boson decay to bottom quarks,
Phys. Rev. Lett. \textbf{121}, 121801 (2018).

\bibitem{CMS:2018uag}
A.~M.~Sirunyan \textit{et al.} [CMS],
Combined measurements of Higgs boson couplings in proton\textendash{}proton collisions at $\sqrt{s}=13\,\text {Te}\text {V} $,
Eur. Phys. J. C \textbf{79}, 421 (2019).

\bibitem{ATLAS:2019nkf}
G.~Aad \textit{et al.} [ATLAS],
Combined measurements of Higgs boson production and decay using up to $80$ fb$^{-1}$ of proton-proton collision data at $\sqrt{s}=$ 13 TeV collected with the ATLAS experiment,
Phys. Rev. D \textbf{101}, 012002 (2020).

\bibitem{Noguchi:2019ofj}
Y.~Noguchi [ATLAS],
Observation of the Higgs decay to beauty quarks,
J. Phys. Conf. Ser. \textbf{1390}, 012046 (2019).

\bibitem{CMS:2017zyp}
A.~M.~Sirunyan \textit{et al.} [CMS],
Observation of the Higgs boson decay to a pair of $\tau$ leptons with the CMS detector,
Phys. Lett. B \textbf{779}, 283-316 (2018).

\bibitem{Cepeda:2019klc}
M.~Cepeda, S.~Gori, P.~Ilten, M.~Kado, F.~Riva, R.~Abdul Khalek, A.~Aboubrahim, J.~Alimena, S.~Alioli and A.~Alves, \textit{et al.}
Report from Working Group 2: Higgs Physics at the HL-LHC and HE-LHC,
CERN Yellow Rep. Monogr. \textbf{7}, 221-584 (2019).

\bibitem{CEPCStudyGroup:2018ghi}
J.~B.~Guimar\~aes da Costa \textit{et al.} [CEPC Study Group],
CEPC Conceptual Design Report: Volume 2 - Physics \& Detector,
arXiv:1811.10545.

\bibitem{ILC:2013jhg}
H.~Baer \textit{et al.} [ILC],
The International Linear Collider Technical Design Report - Volume 2: Physics,
arXiv:1306.6352.

\bibitem{FCC:2018evy}
A.~Abada \textit{et al.} [FCC],
FCC-ee: The Lepton Collider: Future Circular Collider Conceptual Design Report Volume 2,
Eur. Phys. J. ST \textbf{228}, 261-623 (2019).

\bibitem{MuonCollider:2022xlm}
J.~de Blas \textit{et al.} [Muon Collider],
The physics case of a 3 TeV muon collider stage,
arXiv:2203.07261.

\bibitem{Black:2022cth}
K.~M.~Black, S.~Jindariani, D.~Li, F.~Maltoni, P.~Meade, D.~Stratakis, D.~Acosta, R.~Agarwal, K.~Agashe and C.~Aim\`e, \textit{et al.}
Muon Collider Forum Report,
arXiv:2209.01318.

\bibitem{Qiao:1998kv}
C.~F.~Qiao, F.~Yuan and K.~T.~Chao,
Quarkonium production in SM Higgs decays,
J. Phys. G \textbf{24}, 1219-1226 (1998).

\bibitem{Bodwin:2013gca}
G.~T.~Bodwin, F.~Petriello, S.~Stoynev and M.~Velasco,
Higgs boson decays to quarkonia and the $H\bar{c}c$  coupling,
Phys. Rev. D \textbf{88}, 053003 (2013).

\bibitem{Bodwin:2014bpa}
G.~T.~Bodwin, H.~S.~Chung, J.~H.~Ee, J.~Lee and F.~Petriello,
Relativistic corrections to Higgs boson decays to quarkonia,
Phys. Rev. D \textbf{90}, 113010 (2014).

\bibitem{Konig:2015qat}
M.~K\"onig and M.~Neubert,
Exclusive Radiative Higgs Decays as Probes of Light-Quark Yukawa Couplings,
JHEP \textbf{08}, 012 (2015).

\bibitem{Zhou:2016sot}
C.~Zhou, M.~Song, G.~Li, Y.~J.~Zhou and J.~Y.~Guo,
Next-to-leading order QCD corrections to Higgs boson decay to quarkonium plus a photon,
Chin. Phys. C \textbf{40}, 123105 (2016).

\bibitem{Modak:2016cdm}
T.~Modak, J.~C.~Rom\~ao, S.~Sadhukhan, J.~P.~Silva and R.~Srivastava,
Constraining wrong-sign $hbb$ couplings with $h \rightarrow \Upsilon \gamma$,
Phys. Rev. D \textbf{94}, 075017 (2016).

\bibitem{Bodwin:2016edd}
G.~T.~Bodwin, H.~S.~Chung, J.~H.~Ee and J.~Lee,
New approach to the resummation of logarithms in Higgs-boson decays to a vector quarkonium plus a photon,
Phys. Rev. D \textbf{95}, 054018 (2017).

\bibitem{Bodwin:2017wdu}
G.~T.~Bodwin, H.~S.~Chung, J.~H.~Ee and J.~Lee,
Addendum: New approach to the resummation of logarithms in Higgs-boson decays to a vector quarkonium plus a photon [Phys. Rev. D 95, 054018 (2017)],
Phys. Rev. D \textbf{96}, 116014 (2017).

\bibitem{Sun:2018xft}
Q.~F.~Sun and A.~M.~Wang,
Next-to-leading order QCD corrections to the decay of Higgs to vector meson and Z boson,
Chin. Phys. C \textbf{42}, 033105 (2018).

\bibitem{Liao:2018nab}
Q.~L.~Liao, Y.~Deng, Y.~Yu, G.~C.~Wang and G.~Y.~Xie,
Heavy $P$-wave quarkonium production via Higgs decays,
Phys. Rev. D \textbf{98}, 036014 (2018).

\bibitem{Brambilla:2019fmu}
N.~Brambilla, H.~S.~Chung, W.~K.~Lai, V.~Shtabovenko and A.~Vairo,
Order $v^4$ corrections to Higgs boson decay into $J/\psi + \gamma$,
Phys. Rev. D \textbf{100}, 054038 (2019).

\bibitem{Mao:2019hgg}
S.~Mao, Y.~Guo-He, L.~Gang, Z.~Yu and G.~Jian-You,
Probing the charm-Higgs Yukawa coupling via Higgs boson decay to $h_c$ plus a photon,
J. Phys. G \textbf{46}, 105008 (2019).

\bibitem{Liao:2019xux}
Q.~L.~Liao and J.~Jiang,
Excited heavy quarkonium production in Higgs boson decays,
Phys. Rev. D \textbf{100}, 053002 (2019).

\bibitem{Sun:2019cxx}
Z.~Sun and Y.~Ma,
Inclusive productions of $\Upsilon(1S,2S,3S)$ and $\chi_b(1P,2P,3P)$ via the Higgs boson decay,
Phys. Rev. D \textbf{100}, 094019 (2019).

\bibitem{Pan:2022nxc}
X.~A.~Pan, Z.~M.~Niu, M.~Song, Y.~Zhang, G.~Li and J.~Y.~Guo,
J/\ensuremath{\psi} associated production with a bottom quark pair from the Higgs boson decay in next-to-leading order QCD,
Phys. Rev. D \textbf{105}, 014032 (2022).

\bibitem{Han:2022rwq}
T.~Han, A.~K.~Leibovich, Y.~Ma and X.~Z.~Tan,
Higgs boson decay to charmonia via c-quark fragmentation,
JHEP \textbf{08}, 073 (2022).

\bibitem{Gao:2022iam}
D.~N.~Gao and X.~Gong,
Higgs boson decays into a pair of heavy vector quarkonia,
Phys. Lett. B \textbf{832}, 137243 (2022).

\bibitem{Batra:2022wsd}
A.~Batra, S.~Mandal and R.~Srivastava,
$h \to \Upsilon \gamma$ Decay: Smoking Gun Signature of Wrong-Sign $hb\bar{b}$ Coupling,
arXiv:2209.01200.

\bibitem{Jiang:2015pah}
J.~Jiang and C.~F.~Qiao,
$B_c$ Production in Higgs Boson Decays,
Phys. Rev. D \textbf{93}, 054031 (2016).

\bibitem{Braaten:1980yq}
E.~Braaten and J.~P.~Leveille,
Higgs Boson Decay and the Running Mass,
Phys. Rev. D \textbf{22}, 715 (1980).

\bibitem{Sakai:1980fa}
N.~Sakai,
Perturbative QCD Corrections to the Hadronic Decay Width of the Higgs Boson,
Phys. Rev. D \textbf{22}, 2220 (1980).

\bibitem{Zheng:2019egj}
X.~C.~Zheng, C.~H.~Chang, X.~G.~Wu, J.~Zeng and X.~D.~Huang,
Next-to-leading order QCD corrections to the production of $B_c$ and $B_c^*$ through $W^+$-boson decays,
Phys. Rev. D \textbf{101}, 034029 (2020).

\bibitem{Zhang:2021ypo}
Z.~Y.~Zhang, X.~C.~Zheng and X.~G.~Wu,
Production of the $B_c$ meson at the CEPC,
Eur. Phys. J. C \textbf{82}, 246 (2022).

\bibitem{Li:2019tbn}
Q.~Li, M.~S.~Liu, L.~S.~Lu, Q.~F.~L\"u, L.~C.~Gui and X.~H.~Zhong,
Excited bottom-charmed mesons in a nonrelativistic quark model,
Phys. Rev. D \textbf{99}, 096020 (2019).

\bibitem{Gorishnii:1990zu}
S.~G.~Gorishnii, A.~L.~Kataev, S.~A.~Larin and L.~R.~Surguladze,
Corrected Three Loop QCD Correction to the Correlator of the Quark Scalar Currents and $\Gamma_{\rm tot}(H^0 \to {\rm Hadrons})$,
Mod. Phys. Lett. A \textbf{5}, 2703-2712 (1990).

\bibitem{Kataev:1993be}
A.~L.~Kataev and V.~T.~Kim,
The Effects of the QCD corrections to $\Gamma (H^0 \to b \bar{b})$,
Mod. Phys. Lett. A \textbf{9}, 1309-1326 (1994).

\bibitem{Surguladze:1994gc}
L.~R.~Surguladze,
Quark mass effects in fermionic decays of the Higgs boson in O (alpha-s**2) perturbative QCD,
Phys. Lett. B \textbf{341}, 60-72 (1994).

\bibitem{Larin:1995sq}
S.~A.~Larin, T.~van Ritbergen and J.~A.~M.~Vermaseren,
The Large top quark mass expansion for Higgs boson decays into bottom quarks and into gluons,
Phys. Lett. B \textbf{362}, 134-140 (1995).

\bibitem{Chetyrkin:1995pd}
K.~G.~Chetyrkin and A.~Kwiatkowski,
Second order QCD corrections to scalar and pseudoscalar Higgs decays into massive bottom quarks,
Nucl. Phys. B \textbf{461}, 3-18 (1996).

\bibitem{Chetyrkin:1996sr}
K.~G.~Chetyrkin,
Correlator of the quark scalar currents and $\Gamma_{\rm tot} (H \to {\rm hadrons})$ at $O(\alpha_s^3)$ in pQCD,
Phys. Lett. B \textbf{390}, 309-317 (1997).

\bibitem{Baikov:2005rw}
P.~A.~Baikov, K.~G.~Chetyrkin and J.~H.~Kuhn,
Scalar correlator at $O(\alpha_s^4)$, Higgs decay into b-quarks and bounds on the light quark masses,
Phys. Rev. Lett. \textbf{96}, 012003 (2006).

\bibitem{Mondini:2019gid}
R.~Mondini, M.~Schiavi and C.~Williams,
N$^{3}$LO predictions for the decay of the Higgs boson to bottom quarks,
JHEP \textbf{06}, 079 (2019).

\bibitem{Bodwin:1994jh}
G.~T.~Bodwin, E.~Braaten and G.~P.~Lepage,
Rigorous QCD analysis of inclusive annihilation and production of heavy quarkonium,
Phys. Rev. D \textbf{51}, 1125-1171 (1995).

\bibitem{Chang:1992bb}
C.~H.~Chang and Y.~Q.~Chen,
The Production of B(c) or anti-B(c) meson associated with two heavy quark jets in Z0 boson decay,
Phys. Rev. D \textbf{46}, 3845 (1992);
erratum: Phys. Rev. D \textbf{50}, 6013 (1994).

\bibitem{Chang:1991bp}
C.~H.~Chang and Y.~Q.~Chen,
The B(c) and anti-B(c) mesons accessible to experiments through Z0 bosons decay,
Phys. Lett. B \textbf{284}, 127-132 (1992).

\bibitem{Braaten:1993jn}
E.~Braaten, K.~m.~Cheung and T.~C.~Yuan,
Perturbative QCD fragmentation functions for $B_c$ and $B_{c}$ * production,
Phys. Rev. D \textbf{48}, R5049 (1993).

\bibitem{Ma:1994zt}
J.~P.~Ma,
Calculating fragmentation functions from definitions,
Phys. Lett. B \textbf{332}, 398-404 (1994).

\bibitem{Zheng:2019gnb}
X.~C.~Zheng, C.~H.~Chang, T.~F.~Feng and X.~G.~Wu,
QCD NLO fragmentation functions for c or $\bar{b}$ quark to $B_c$ or $B_c^*$ meson and their application,
Phys. Rev. D \textbf{100}, 034004 (2019).

\bibitem{Zheng:2021sdo}
X.~C.~Zheng, C.~H.~Chang and X.~G.~Wu,
Fragmentation functions for gluon into $B_c$ or $B_c^{*}$ meson,
JHEP \textbf{05}, 036 (2022).

\bibitem{Feng:2021qjm}
F.~Feng, Y.~Jia and D.~Yang,
Gluon fragmentation into $Bc(*)$ in NRQCD factorization,
Phys. Rev. D \textbf{106}, 054030 (2022).

\bibitem{Curci:1980uw}
G.~Curci, W.~Furmanski and R.~Petronzio,
Evolution of Parton Densities Beyond Leading Order: The Nonsinglet Case,
Nucl. Phys. B \textbf{175}, 27-92 (1980).

\bibitem{Furmanski:1980cm}
W.~Furmanski and R.~Petronzio,
Singlet Parton Densities Beyond Leading Order,
Phys. Lett. B \textbf{97}, 437-442 (1980).

\bibitem{Floratos:1978ny}
E.~G.~Floratos, D.~A.~Ross and C.~T.~Sachrajda,
Higher Order Effects in Asymptotically Free Gauge Theories. 2. Flavor Singlet Wilson Operators and Coefficient Functions,
Nucl. Phys. B \textbf{152}, 493-520 (1979).

\bibitem{Gonzalez-Arroyo:1979qht}
A.~Gonzalez-Arroyo and C.~Lopez,
Second Order Contributions to the Structure Functions in Deep Inelastic Scattering. 3. The Singlet Case,
Nucl. Phys. B \textbf{166}, 429-459 (1980).

\bibitem{Floratos:1981hs}
E.~G.~Floratos, C.~Kounnas and R.~Lacaze,
Higher Order QCD Effects in Inclusive Annihilation and Deep Inelastic Scattering,
Nucl. Phys. B \textbf{192}, 417-462 (1981).

\bibitem{Hirai:2011si}
M.~Hirai and S.~Kumano,
Numerical solution of $Q^2$ evolution equations for fragmentation functions,
Comput. Phys. Commun. \textbf{183}, 1002-1013 (2012).

\bibitem{ParticleDataGroup:2022pth}
R.~L.~Workman \textit{et al.} [Particle Data Group],
Review of Particle Physics,
PTEP \textbf{2022}, 083C01 (2022).

\bibitem{Eichten:1995ch}
E.~J.~Eichten and C.~Quigg,
Quarkonium wave functions at the origin,
Phys. Rev. D \textbf{52}, 1726-1728 (1995).

\bibitem{Chetyrkin:1997dh}
K.~G.~Chetyrkin,
Quark mass anomalous dimension to O (alpha-s**4),
Phys. Lett. B \textbf{404}, 161-165 (1997).

\bibitem{Vermaseren:1997fq}
J.~A.~M.~Vermaseren, S.~A.~Larin and T.~van Ritbergen,
The four loop quark mass anomalous dimension and the invariant quark mass,
Phys. Lett. B \textbf{405}, 327-333 (1997).

\bibitem{Herren:2017osy}
F.~Herren and M.~Steinhauser,
Version 3 of RunDec and CRunDec,
Comput. Phys. Commun. \textbf{224}, 333-345 (2018).

\bibitem{Gray:1990yh}
N.~Gray, D.~J.~Broadhurst, W.~Grafe and K.~Schilcher,
Three Loop Relation of Quark (Modified) Ms and Pole Masses,
Z. Phys. C \textbf{48}, 673-680 (1990).

\bibitem{Broadhurst:1991fy}
D.~J.~Broadhurst, N.~Gray and K.~Schilcher,
Gauge invariant on-shell Z(2) in QED, QCD and the effective field theory of a static quark,
Z. Phys. C \textbf{52}, 111-122 (1991).

\bibitem{Chetyrkin:1999ys}
K.~G.~Chetyrkin and M.~Steinhauser,
Short distance mass of a heavy quark at order $\alpha_s^3$,
Phys. Rev. Lett. \textbf{83}, 4001-4004 (1999).

\bibitem{Melnikov:2000qh}
K.~Melnikov and T.~v.~Ritbergen,
The Three loop relation between the MS-bar and the pole quark masses,
Phys. Lett. B \textbf{482}, 99-108 (2000).

\bibitem{Marquard:2015qpa}
P.~Marquard, A.~V.~Smirnov, V.~A.~Smirnov and M.~Steinhauser,
Quark Mass Relations to Four-Loop Order in Perturbative QCD,
Phys. Rev. Lett. \textbf{114}, 142002 (2015).

\bibitem{Hahn:2000kx}
T.~Hahn,
Generating Feynman diagrams and amplitudes with FeynArts 3,
Comput. Phys. Commun. \textbf{140}, 418-431 (2001).

\bibitem{Mertig:1990an}
R.~Mertig, M.~Bohm and A.~Denner,
FEYN CALC: Computer algebraic calculation of Feynman amplitudes,
Comput. Phys. Commun. \textbf{64}, 345-359 (1991).

\bibitem{Shtabovenko:2016sxi}
V.~Shtabovenko, R.~Mertig and F.~Orellana,
New Developments in FeynCalc 9.0,
Comput. Phys. Commun. \textbf{207}, 432-444 (2016).

\bibitem{Chang:1996jt}
C.~H.~Chang, Y.~Q.~Chen and R.~J.~Oakes,
Comparative study of the hadronic production of B(c) mesons,
Phys. Rev. D \textbf{54}, 4344-4348 (1996).


\bibitem{Eichten:2019gig}
E.~J.~Eichten and C.~Quigg,
Mesons with Beauty and Charm: New Horizons in Spectroscopy,
Phys. Rev. D \textbf{99}, 054025 (2019).




\end{thebibliography}
\end{document}